\documentclass[12pt]{article}
\usepackage{hyperref}
\hypersetup{
colorlinks=true,
linkcolor=black,
citecolor=black,
urlcolor=cyan,
filecolor=black,
pdfborder={0 0 0},
}

\usepackage{amssymb}
\usepackage{amsfonts}
\usepackage{mathrsfs}
\usepackage{latexsym}
\usepackage{amsmath}
\usepackage{graphics}
\usepackage{graphicx}
\usepackage{sectsty}
\usepackage{setspace}
\usepackage{multirow}
\usepackage{cite}
\sectionfont{\normalsize\bf}
\subsectionfont{\rm\normalsize}
\def\be{\begin{equation}}
\def\ee{\end{equation}}
\def\bd{\left|\begin{matrix}}
\def\ed{\end{matrix}\right|}
\input amssym.def
\input amssym
\def\Black{} 

\begin{document}

\thispagestyle{empty}

\baselineskip=16pt
\vspace{.5in}
{
\begin{center}
{\bf The Guiding Influence of Stanley Mandelstam,}

\bigskip

{\bf from $\boldsymbol S$-Matrix Theory to String Theory}

\end{center}}
\vskip 1.1cm
\begin{center}

\bigskip
\bigskip        
{Peter Goddard}
\vskip15pt

\centerline{\em School of Natural Sciences}
\centerline{\em Institute for Advanced Study}
\centerline{\em Princeton, NJ 08540, USA}
\bigskip
\bigskip
\bigskip
\bigskip
\end{center}
\abstract{\noindent 
The guiding influence of some of Stanley Mandelstam's key contributions to the development of theoretical high energy physics is discussed, from the motivation for the study of the analytic properties of the scattering matrix through to dual resonance models and their 
evolution into string theory.

  }
  
\bigskip
\vfill
\centerline{a contribution to {\it Memorial Volume for Stanley Mandelstam,}}
\centerline{ed. N. Berkovits {\it et al.} (World Scientific, Singapore, 2017) }

\setlength{\parindent}{0pt}
\setlength{\parskip}{6pt}

\setstretch{1.05}
\eject
\vskip50pt

\section{The Mandelstam Representation}

When I began research on the theory of the strong interactions in Cambridge in 1967, the focuses of study were the Regge theory of the high energy behavior of scattering amplitudes, and the properties of these amplitudes as analytic functions of complex variables. Most prominent amongst the names conjured with in these subjects was that of Stanley Mandelstam: the complex variables that the scattering amplitudes depended on were the {\it Mandelstam variables}; the complex space they varied over was the {\it Mandelstam diagram}; and the proposal of the {\it Mandelstam representation} had provided the inspiration for much of the study of the analytic properties of scattering amplitudes that was then in full spate. 

Entering the field then, one was not readily conscious of the fact that this conceptual framework had its origins less than ten years earlier, in 1958--59, in a seminal series of papers \cite{M1,M2,M3} by Mandelstam. The book, {\it The Analytic S-Matrix}, by Eden, Landshoff, Olive and Polkinghorne \cite{ELOP}, published in 1966, the bible for research students in Cambridge at the time, begins with the slightly arch sentence, ``One of the most important discoveries in elementary particle physics has been that of the complex plane''. The ideas of analyticity in energy had been around for sometime, used, for example, to derive dispersion relations, essentially by application of Cauchy's theorem (as reviewed by Mandelstam in \cite{M4}). The real breakthrough that Mandelstam made, which underlay and motivated the developments described in that still relevant book, was to show how scattering amplitudes could and should be thought of as functions of {\it more than one} complex variable. 

Like many deep insights, Mandelstam's perception that scattering amplitudes could be regarded as analytic functions of  momentum invariants, although so absorbed into the conceptual framework of particle physics that it is taken for granted now, initially was difficult for some to accept, as Geoffrey Chew  recounts \cite{GFC1}. Marvin Goldberger listed among his excuses for failing to understand the significance of what Mandelstam was doing 
at that time that he had ``never understood a word that Stanley says on any subject. He is  almost always right, has fantastic understanding, intuition, and mathematical power, but to me he is far from lucid in his presentation of his wisdom'' \cite{MLG}. Certainly, unwittingly on his part, Mandelstam's formidable  technical skills could make his papers a steep uphill journey for others who were less gifted in this regard.

Mandelstam \cite{M1} considered the two-to-two scattering amplitude, $A$, as an analytic function of $s$ and $t$, the square of center of mass momentum and the momentum transfer invariant, respectively; so these and, more generally, the square of the sum of any subset of the momenta in an $N$-particle scattering process became known as the Mandelstam variables. (See \cite{CBT} for an excellent detailed account of Mandelstam's research in the period discussed in this article.) 

Viewing the scattering amplitude for pion-nucleon scattering, $A(s,t)$, as an analytic function of $s$ and $t$, with assumptions about the singularity structure of $A(s,t)$, which could be established in low order perturbation theory in quantum field theory (QFT), Mandelstam could apply Cauchy's theorem twice to obtain a `double dispersion relation', and thus established a new representation of the scattering amplitude, which became known as the {\it Mandelstam representation}. Although Chew was skeptical at first, Mandelstam quickly convinced him that  this was the right way to think of the two-to-two scattering amplitude, effectively showing how to analytically continue  the amplitude in energy and angle, something that Chew had in fact been trying to do for some time \cite{FC}. 

Chew immediately recruited Mandelstam to Berkeley, and, apart from a period 1960-63 as professor back in Birmingham, where he had taken his PhD in the group of Rudolf Peierls, he remained in Berkeley for the rest of his life. Before graduate studies in Birmingham, and following his first degree, a BSc from  Witwatersrand, Mandelstam took a BA from Trinity College, Cambridge in 1954, sharing the university prize for the best performance in applied mathematics and theoretical physics  in the final degree examination with Jeffrey Goldstone, also at Trinity. At that time, Paul Dirac held the Lucasian Professorship in Cambridge, as he still did when I was a graduate student. When, in 1967, a chair assigned to theoretical particle physics was established, I believe an attempt was made to attract Mandelstam back to Cambridge, but he could not be moved from Berkeley.

\section{$\boldsymbol S$-Matrix Theory}

Mandelstam and Chew began collaborating, applying Mandelstam's approach to the study of pion-pion scattering  \cite{CM1,CM2}. In this context, the idea of `crossing',  whereby the analytic continuation of the amplitude $A(s,t)$ for the process $A+B\rightarrow C+D$, from the physical region, in which $s>0,\, t<0$, to a   region in which $s<0,\, t>0$, gives the amplitude for the process $A+\overline C\rightarrow \overline B+D$, where $\overline B$ is the antiparticle of $B$,
could actually be used in dynamical calculations, relating bound states in one channel  to forces in the other. 
The development of these ideas led Chew to formulate his `bootstrap hypothesis'. This was the proposition that the requirements of analyticity, which was related to causality, and the requirement of unitarity, which seemed fundamental to the probabilistic interpretation of quantum theory, perhaps together with some suitable asymptotic assumptions at high energy, should  determine uniquely the scattering amplitudes, {\it i.e.} the $S$-matrix. 
Along with this went Chew's principle of `nuclear democracy', in which no particles would be more fundamental than any other but all were in some sense bound states of one another. 

Mandelstam's calculations in perturbative QFT showed that the amplitudes had poles and branch points. That these singularities are present could be seen to be a necessary consequence of unitarity. 
Omn\`es and Froissart, in their 1963 text book {\it Mandelstam Theory and Regge Poles} \cite{OF}, set out what they called  the assumption that the scattering amplitude only have the singularities required by unitarity `The Mandelstam  Hypothesis'. The hope of Chew's bootstrap approach, a hope that became almost a tenet of faith, was that the scattering amplitudes would be determined uniquely if this `Mandelstam  Hypothesis' held.

 Chew was a powerful and charismatic  evangelist for Mandelstam's conceptual framework. 
Polkinghorne recalled it being said in the early 1960s, with a different religious metaphor,  that  ``There is no God but Mandelstam and Chew is his prophet'' \cite{JCP}. But it seems that, in this context, God himself was an agnostic, for Mandelstam never fully subscribed to the bootstrap. As Chew put it, ``Stanley never endorsed it but, being a very mild person, he did not fight it, and the term also appeared in one of our papers. ...
He feels a need for something fundamental ... he always believed he had firm ground under his feet''\cite{FC}. Reading Mandelstam's papers one certainly feels the firm ground under one's feet, and for him at that time, firm ground meant calculations based on perturbative QFT.

At the beginning of the 1960s, Mandelstam's work presented two great theoretical challenges, different in direction: the first, from the point of view of QFT, was to prove the Mandelstam representation, order by order in perturbation theory; and, the second, from the $S$-matrix point of view, was to determine the singularity structure of the scattering amplitudes implied by unitarity. Both of these challenges are discussed in the book of Eden {\it et al.} \cite{ELOP}. In 1961, Landshoff, Polkinghorne and Taylor \cite{LPT} and, independently, Eden \cite{RJE} published proofs of the Mandelstam representation in perturbation theory, but, unfortunately, later that year the four authors found that their arguments were incomplete, because there are terms in perturbation theory that possess isolated real singularities,  anodes, to which are attached complex singularities which vitiate the proof of the Mandelstam representation \cite{ELPT}.  While, a proof is still lacking,  these investigations did lead to a deeper understanding of the singularities of Feynman integrals.  

The program of determining the singularity structure of the $S$-matrix from unitarity was carried forward most extensively by David Olive, as described in chapter 4 of \cite{ELOP}, and in the book, {\it The Analytic S Matrix},
by Chew \cite{GFC2}, published in 1966 at almost exactly the same time as \cite{ELOP} (but lacking the hyphen), which relies heavily on Olive's work in its early chapters.  Later, Chew saw this book as presenting the culmination of his bootstrap ideas, but with some disappointment because, in trying to make as complete as possible a dynamical analysis of pion-pion system within this framework, they had not been able to go beyond two-particle branch cuts in the analytic structure of the scattering amplitudes, in order to take account of singularities corresponding to multi-particle states \cite{FC}.

With colleagues in Cambridge, Olive developed his methods further  to describe in some generality the singularities of the $S$-matrix at real points of the physical region of the Mandelstam variables and the corresponding discontinuities \cite{BOP}. However, little progress has been made in describing singularities, real or complex, outside the physical region, which would be necessary in order to discuss the validity of the Mandelstam representation. 

\section{Regge Theory}

As Chew explains in his book, for the bootstrap principle conceivably to lead to a unique possible $S$-matrix, some asymptotic conditions on the amplitudes  for large values of $s, t,$ {\it etc.,} must be imposed, just as for an analytic function of a single  complex variable, $z$, to be constrained severely, an asymptotic condition on the function is required, such as being bounded  by a power of $z$ for $|z|\rightarrow\infty$. Fortuitously, in 1959, just before Chew and Mandelstam's study of pion-pion scattering,  work of Tullio Regge \cite{R1,R2}, aimed at proving the Mandelstam representation for potential scattering, led to developments that provided what just was needed here and, indeed, reshaped much of theoretical research on strong interaction physics throughout the 1960s. 

Regge showed that the amplitude, $a_J$, for  the non-relativistic scattering of a particle off a potential at a specific angular momentum, $J$, could be continued away from integer $J$ to complex values, so as to yield an analytic function of $J$, and this might have poles at $J=\alpha_E$, depending on the energy, $E$; these became known as Regge poles. For a given angular momentum, $J$, $J=\alpha_E$ determines the energy at which a bound state or resonance occurs. Thus, the curve, $J=\alpha_E$, known as a Regge trajectory, relates resonances with different angular momenta, or spin, corresponding to different integer values.
If one considers analytically continuing the scattering amplitude, $a(E,\theta)$, at scattering angle $\theta$ for fixed energy, $E$, to the unphysical region of large $z=\cos\theta$, it behaves like $z^{\alpha_E}$ as $z\rightarrow\infty$, where $\alpha_E$ is the Regge trajectory for which the real part of $\alpha_E$ is largest for the given value of $E$, called the leading Regge trajectory. [Here, for simplicity, we shall just refer to meson resonances with integer spin.]

It was Mandelstam who first argued for the importance of Regge theory for high energy particle physics, through unpublished discussions with Chew, Frautschi and others, for which he never sought credit \cite{SCF1,SCF2}. He saw that Regge theory applied to relativistic scattering would provide the appropriate boundary condition for the $S$-matrix \cite{SCF3}. As Elliot Leader later put it,
Regge's great imaginative leap, of introducing complex angular momentum in non-relativistic quantum mechanics, might have ended in oblivion if Mandelstam had not  demonstrated its striking consequences in high-energy scattering processes \cite{EL} (quoted in \cite{DR}). 

In the relativistic context,  the energy $E$ can be replaced by $s$ and $z=\cos\theta$ by $t$, at fixed $s$. Then a Regge trajectory takes the form $J=\alpha(s)$, and corresponds to a sequence of resonances at mass squared $s$ given by integral $J$, and determines a contribution $\beta(s) t^{\alpha(s)}$ to the behavior of the amplitude in the limit $t\rightarrow\infty$ at fixed $s$. The difference in the relativistic case is that this unphysical limit for the original process, $A+B\rightarrow C+D$ (the $s$-channel), is a physical limit, the high energy limit, for the `crossed channel' $A+\overline C\rightarrow \overline B+D$ (the $t$-channel). 
Resonances in one channel are related to high energy behavior in the crossed channel. 

Stimulated by Mandelstam, Chew and Frautschi \cite{CF} proposed that the known strongly interacting particles  lay on nearly straight and parallel Regge trajectories and then Mandelstam collaborated with them on developing a calculational procedure
for analyzing the interrelation between Regge poles, bound states, resonances and high energy behavior, arguing that Regge poles and behavior were general phenomena for strongly interacting particles \cite{CFM}. 

In his book \cite{GFC2}, Chew supplemented the assumption that the $S$-matrix should be as analytic as possible in the Mandelstam variables, consistent with unitarity, with an assumption of suitable analyticity in $J$, including that all particles should lie on Regge trajectories, and allowing for the possibility that there might be singularities other than poles in the complex $J$ plane. 
In 1962, Amati, Fubini and Stanghellini (AFS) \cite{AFS} proposed a mechanism to show that there should be cuts in the $J$ plane based on  perturbative QFT, but, soon after, Mandelstam \cite{M5} demonstrated both that the  AFS cuts were in fact cancelled by other contributions and that an adaptation of their mechanism would produce a cut that would actually be present in the $J$ plane. In so doing, he took the major step in establishing the nature of the singularity structure in the complex angular momentum plane.

\section{Dual Models and  String Theory}

While a great deal of effort in strong interaction phenomenology from the mid 1960s onwards was devoted to understanding scattering data in terms of sums of Regge pole, and, where necessary, Regge cut, contributions at high energy and in terms of sums of resonance contributions at lower energies, the question arose as to whether these two sorts of description should be added together to get a more accurate parameterization of the scattering amplitudes or whether they should be regarded as alternative descriptions of the data, equivalent, or dual,  in some sense. The former procedure, in which Regge contributions should be added to resonance contributions, was known as `interference', and the latter, in which they were equivalent to one another, and which the description at high energy in terms of Regge poles could be determined from the parametrization of lower energy data in terms of resonances \cite{DHS}, was known as `duality'.

For a time, it seemed difficult to see analytically how a description of the scattering amplitude as a sum of  resonance poles could be equivalent to an asymptotic expansion in terms of Regge poles, and there were suspicions that it might be mathematically impossible for this to be exactly the case, until, in the summer of 1968, Gabriele Veneziano produced, almost as a sort  of {\it deus ex machina}, what very quickly became his famous eponymous formula \cite{GV}, $A(s,t)=B(-\alpha(s),-\alpha(t))$, where $B$ is the classical Euler Beta function and the Regge trajectory is linear, $\alpha(s)=\alpha_0+\alpha' s$. This is symmetric in $s$ and $t$, meromorphic in $s$ with poles whose residues are polynomials in $t$ (and {\it vice versa}); it can be written as a sum over these poles or, equivalently, an asymptotic series of Regge pole contributions, realizing duality in a mathematical precise form. The resonances in one channel are precisely dual to the Regge poles in the other.

The Veneziano formula, describing two-to-two scattering, was generalized quickly to processes involving an arbitrary number of particles, by the construction of $N$-particle amplitudes which were also meromorphic and possessing Regge behavior, in an appropriate sense, for large values of Mandelstam variables. It was also generalized by finding other formulae for two particle scattering amplitudes, which possessed similar properties to the Veneziano formula. At first, these proposals, known as dual models, were either directed at providing sufficient flexibility in the amplitudes to enable them to be used to fit experimental data or to provide theoretical laboratories within which to study the form Regge theory might take for multi-particle processes. 

However, soon it was appreciated that the Veneziano amplitude could be taken more seriously, in a sense, as being the starting point for a perturbative expansion for the scattering amplitudes, consistent with unitarity at least as a formal power series, in a similar way to perturbative QFT. Indeed, $S$-matrix theory, which in Chew's formulation of the bootstrap had offered the hope of determining the $S$-matrix uniquely, could also more modestly and more concretely provide the conceptual framework within which dual models could be viewed on comparable terms with perturbative QFT, as providing a model theory formally satisfying the basic postulates.
For Mandelstam, who had been reluctant to give up the crutch  that (perturbative) QFT provided to $S$-matrix theory, this gave an appealing alternative way to keep one's feet on solid ground. He later commented, after dual models had metamorphosed into string theory, ``The string model originated as a model for the $S$-matrix, and it may well not have been discovered if $S$-matrix theory had not been vigorously pursued at the time'' \cite{M6}.

The Veneziano amplitude, extended to $N$-particle processes, is the starting point for a perturbative expansion for an $S$-matrix; higher order contributions need to be added, containing threshold cuts and the other singularities, corresponding to loop contributions in QFT. But, in order for it to be an appropriate starting point, even before considering whether the higher order contributions can be defined consistently, there are conditions that the $N$-particle Veneziano amplitude needs to satisfy, as a consequence of unitarity. In particular, the residue at each pole in a Mandelstam variable needs to factorize in the sense of being a sum over intermediate states of products of two amplitudes for fewer particles. In fact, this process, if it can be done consistently, determines the particle spectrum of the theory. 

Mandelstam's early work had set much of the stage for the introduction of dual models, and, more recently, he had provided some of the immediate motivation \cite{M14} for Veneziano's breakthrough.
He was quick to make a fundamental contribution to their development by establishing, with Bardacki \cite{BM} the factorization of the generalized Veneziano model, simultaneously with work of Fubini and Veneziano \cite{FV}, just nine months after that breakthrough. Their results demonstrated an unanticipated exponential growth in the degeneracy of states with mass, typical of a vibrating extended one-dimensional medium, as was pointed out over the following year by Nambu, Nielsen and Susskind   \cite{YN1, HBN, LS}, describing it variously as a rubber band or string. At first, the main impact of this description was through the analogue approach introduced by Nielsen \cite{HBN}, because it suggested appropriate mathematical techniques for  calculating loop contributions, but it was unclear at the time whether it should be regarded merely as an analogy and calculational guide or as a deeper physical description.

The essential aspect of a relativistic string theory, absent in those earliest  analogies, that enables the string to be more than only a qualitative correspondence, is that its (dynamically significant) oscillations should only be transverse. This requires the action for the string to depend just on the surface it traces out in space-time, rather than the particular way it is parametrized. In 1970, Nambu \cite{YN2}, and then Goto \cite{TG}, wrote down such a  reparametrization invariant action. After the structure of the physical states of the dual model, implied by factorization, was completely understood \cite{RCB, GT} in 1972, the way to quantize systematically the Nambu-Goto action, using light cone coordinates \cite{GGRT} became clear.

This showed definitively that the spectrum of the Veneziano dual model was precisely that of the quantized Nambu-Goto string, but it remained to show that the scattering amplitudes, defining the model, actually followed from the Nambu-Goto action. At the time, this seemed to present a formidable technical challenge, but, within nine months, developing a path integral approach, Mandelstam established that the scattering amplitudes for the Nambu-Goto string, as specified directly by their action, integrated over histories in which strings are allowed to split and join, are precisely the Veneziano model amplitudes \cite{M7}. In a series of papers \cite{M7,M8,M9,M10}, he demonstrated the great power of path integral techniques for calculating amplitudes in string theory, although he remained largely alone in using them until Polyakov's seminal work in 1981 \cite{AMP}. In particular, Mandelstam was immediately able to calculate the amplitude for fermion-fermion scattering in the dual model of Neveu, Schwarz and Ramond \cite{NS, PR}, a task that had proved much more difficult using previous techniques \cite{SW, CGOS}. (For a more detailed account see \cite{CBT} and Mandelstam's own recollections \cite{M11}.)

By this time, 1974, string theory had emerged from its origin in dual models, a metamorphosis in which Mandelstam had played a key role. He wrote a review article \cite{M12}, which provided a definitive account of the theory at this stage of its development, before his own interest shifted away from string theory for some years. 

\section{Epilogue}

I first met Stanley Mandelstam when he visited CERN in July 1971, and gave a seminar, entitled {\it Dual resonance models with quarks}, on his program to build a dual model including quarks as a basis for describing hadrons \cite{M13}. Although more than forty-five years ago, I remember clearly meeting someone who had previously seemed almost mythical to me. The impression that was created by his combination of quiet modesty, formidable intellectual strength and lively wit, and his approachability did not diminish at all his status as one of my personal heroes.  

I had come to CERN the previous year as a postdoctoral fellow, after completing my doctorate in Cambridge on $S$-matrix analyticity and Regge theory. I had intended to carry on with this work but, inspired by lectures by David Olive, just before I left Cambridge, and attracted by the lively group of young theoreticians  at CERN working on dual models, which had begun to cluster around Daniele Amati, my interest had shifted to that area. I  felt, as Mandelstam  later said \cite{M6},  that dual models could best be understood at the time as providing an alternative to perturbative QFT as a model for the $S$-matrix, and one that had the advantage for describing strong interactions of possessing Regge behavior at finite order in the perturbative expansion. 

Mandelstam, along with Amati, Fubini, Nambu and (quietly) Goldstone,  was one of the relatively few senior figures in theoretical physics then working on the theory of dual models and lending it their support. The disapproval of many leading physicists, at CERN and elsewhere, gave the research in the area the extra frisson of forbidden fruit. At least in retrospect, ``to be young was very heaven'' then, but the difficulty of getting a permanent post if one worked on string theory was a large part of the reason that interest in the subject largely diminished for a decade after 1974 (see contributions in \cite{CCCV} for accounts of the period). 

In his writings, Stanley Mandelstam was as he was in person: he spoke when he had something to say, and then one should listen. He advanced our understanding in many areas, including: $S$-matrix theory; Regge theory; dual models and string theory; and solitons and monopoles in QFT, always keeping his feet on the ground, making contributions that were both solid and seminal. Perhaps, if he had had less formidable technical skills, some of his work might have been more easily accessible, but the steep climb was easily justified by the wider and deeper view that one obtained.
Although he wrote less than 50 original articles, and would not have stood out exceptionally by the contemporary dull citation metrics, which are now widely and lazily used, his influence remains profound, pervasive and enduring. 
\vskip20pt
\singlespacing


\providecommand{\bysame}{\leavevmode\hbox to3em{\hrulefill}\thinspace}
\providecommand{\MR}{\relax\ifhmode\unskip\space\fi MR }
\providecommand{\MRhref}[2]
{
}
\providecommand{\href}[2]{#2}

\end{document}